\documentclass[12pt]{iopart}

\usepackage{iopams}
\usepackage{bm}
\usepackage{citesort}
\usepackage{graphicx}
\usepackage{graphics,epsfig}
\usepackage{rotating}

\graphicspath{{./}{./EPS/}}

\newcommand{\be}[1]{\begin{equation} \label{(#1)}}
\newcommand{\ee}{\end{equation}}
\newcommand{\baq}[1]{\begin{eqnarray} \label{(#1)}}
\newcommand{\eaq}{\end{eqnarray}}
\newcommand{\rf}[1]{(\ref{(#1)})}

\newcommand*{\vek}[1]{{\ensuremath{\bm{\mathrm{#1}}}}}

\newcommand{\gb}{\bar{\gamma}}

\begin{document}

\title{Static polarizability of two-dimensional hole gases}

\author{Thomas Kernreiter$^{1}$, Michele Governale$^1$ and Ulrich Z\"ulicke$^{2,3}$}

\address{$^1$School of Chemical and Physical Sciences and MacDiarmid Institute for
Advanced Materials and   Nanotechnology, Victoria University of Wellington, PO Box 600, 
Wellington 6140, New Zealand}

\address{$^2$Institute of Fundamental Sciences and MacDiarmid Institute
for Advanced Materials and Nanotechnology, Massey University, Manawatu
Campus, Private~Bag~11~222, Palmerston North 4442, New Zealand}

\address{$^3$Centre for Theoretical Chemistry and Physics, Massey
University, Albany Campus, Private Bag 102904, North Shore MSC,
Auckland 0745, New Zealand}

\eads{\mailto{thomas.kernreiter@vuw.ac.nz}, \mailto{michele.governale@vuw.ac.nz}, \mailto{u.zuelicke@massey.ac.nz}}

\begin{abstract}
We have calculated the density-density (Lindhard) response function of a homogeneous
two-dimensional (2D) hole gas in the static ($\omega=0$) limit. The bulk valence-band 
structure comprising heavy-hole (HH) and light-hole (LH) states is modeled using
Luttinger's $\vek{k}\cdot\vek{p}$ approach within the axial approximation. We elucidate
how, in contrast to the case of conduction electrons, the Lindhard function of 2D holes
exhibits unique features associated with (i)~the confinement-induced HH-LH energy
splitting and (ii)~the HH-LH mixing arising from the charge carriers' in-plane motion.
Implications for the dielectric response and related physical observables are discussed.
\end{abstract}

\pacs{71.10.-w, 71.70.Ej, 73.21.Fg}
\submitto{\NJP}

\maketitle

%%%%%%%%%%%%%%%%%%%%%%%%%%%%%%%%%%%%%%%%%%
\section{Introduction}
%%%%%%%%%%%%%%%%%%%%%%%%%%%%%%%%%%%%%%%%%%

The density-density response function is a very fundamental materials characteristics,
as it determines a host of thermodynamic and transport properties in condensed-matter
systems~\cite{Ziman}. It has been discussed extensively within the paradigmatic model
of the homogeneous electron gas~\cite{GiulianiVignale} and studied for low-dimensional
conductors realised in semiconductor heterostructures~\cite{hetero1}. More recently,
static and dynamic response properties of two-dimensional (2D) conduction-electron
systems with spin-orbit coupling have been investigated in considerable
detail~\cite{raikh:prb:99,wang:prb:05,mishch:prl:06,pletyu:prb:06,fabian:prb:09,qmc}. This
surge of interest arose partly because of important ramifications for possible spintronics
applications~\cite{zutic:rmp:04,flatte:natphys:06}. In contrast, very few studies have
considered how the peculiar electronic properties of a typical semiconductor's valence
band~\cite{YuCardona} affect the polarizability and other many-body response functions
of \textit{p}-type semiconductor materials, and these existing
works~\cite{schlie:prb:06b,galitski:prb:06,schlie:unpub:10} have focused on bulk (3D)
systems. As high-quality 2D holes gases have recently become available for experimental
study, both in
modulation-doped~\cite{ensslin:apl:04,manfra:apl:05,matt:apl:05,weg:apl:05a,weg:apl:05b}
and accumulation-layer~\cite{clarke:jap:06} heterostructures, a detailed theoretical analysis
of their many-body response properties is warranted. Here we provide such a study
and show how the intricate interplay between quantum confinement and strong
spin-orbit-coupled dynamics in the valence band~\cite{rolandbook} has a profound effect
on the static polarizability.

Charge carriers from the conduction and valence bands of typical semiconductors exhibit
profoundly different spin properties. Conduction electrons are quite ordinary in that they
are spin-1/2 particles carrying a fixed intrinsic magnetic dipole moment, like free electrons
in vacuum. Holes are different; they have an intrinsic spin-3/2 degree of freedom because
valence-band states are strongly modified by spin-orbit coupling~\cite{YuCardona}. As a
result, the orbital dynamics of holes in a bulk sample also depends on the magnitude of
projection for their spin parallel to their direction of motion. States with spin-3/2 projection
quantum number $m_J=\pm 3/2$ ($\pm 1/2$) are called heavy holes, HHs (light holes,
LHs), because their band-energy dispersion has a smaller (larger) curvature. When holes
are confined in a 2D heterostructure, the quantum-well growth direction is the natural
spin-quantisation axis (taken to be the $z$ direction in the following), and the difference
in effective masses translates into an energy splitting between the HH and LH subband
edges corresponding to the same transverse orbital bound state~\cite{bastardrev}. As
the in-plane motion couples HH and LH states, 2D holes with finite wave vector
$\vek{k}_\parallel = (k_x, k_y)$ are no longer of purely HH or LH
type~\cite{bastardrev,rolandbook}. While the HH-LH (subband-edge) splitting is easily
accounted for and usually included in theoretical analyses, the HH-LH \emph{mixing\/}
has sometimes been ignored. It may be tempting to make such a simplification, given
that the density of typical 2D hole gases is quite often low enough that only the lowest
(HH-like) subband is occupied. However, detailed analysis shows this approach to be
too crude for most relevant situations~\cite{rolandbook}. Even only qualitatively accurate
predictions basically always require the inclusion of HH-LH mixing alongside the HH-LH
splitting. As we will show below, the density-density response of 2D hole gases is strongly
affected by HH-LH mixing, i.e., is not simply the sum of the response functions of
independent 2D (HH and LH) gases.

This article is organised as follows. In Section~\ref{sec:model}, we introduce our model
for the upper-most valence band of typical semiconductors, which is based on the
Luttinger Hamiltonian~\cite{luttham1,luttham2} in axial
approximation~\cite{hensel:prb:74,treb:prb:79,rolandbook}. The definition and basic
calculational details for the density-density (Lindhard) response function are given in
Sec.~\ref{sec:lindhard}, including analytical results pertaining to the 2D hole gas in
certain limits. We present plots of the numerically determined static polarizability  in
Sec.~\ref{sec:results} and discuss basic features. Our conclusions are given in
Sec.~\ref{sec:concl}.

%%%%%%%%%%%%%%%%%%%%%%%%%%%%%%%%%%%%%%%%%%
\section{Luttinger-model description of a 2D hole system}
\label{sec:model}
%%%%%%%%%%%%%%%%%%%%%%%%%%%%%%%%%%%%%%%%%%

The Luttinger model~\cite{luttham2} provides a useful description of the upper-most
valence band of typical semiconductors in situations where its couplings to the
conduction band and split-off valence band are irrelevant. We adopt this model here
to investigate how the many-body physics of 2D holes is affected by their peculiar
spin-3/2 properties. In principle, more extended~\cite{treb:prb:79,roessler:prb:91}
multiband Hamiltonians could be employed to improve the accuracy of quantitative
predictions. However, to illustrate the qualitatively new features exhibited by 2D hole
gases in contrast to their conduction-electron counterparts, the Luttinger-model
description is adequate. The particular geometry for our case of interest suggests
using, as our starting point, the Luttinger-model Hamiltonian $H_{\mathrm L}$ in
\emph{axial\/} approximation~\cite{hensel:prb:74,treb:prb:79,rolandbook}:
\numparts
\baq{eq:HL} 
H_{\mathrm L} &=& H_0 + H_1 + H_2 \quad , \\
H_0 &=& -\frac{\hbar^2}{2m_0} \left[ \gamma_1 \left( \vek{k}_\parallel^2 + k_z^2
\right) + \tilde\gamma_1 \left( \vek{k}_\parallel^2 - 2 k_z^2 \right) \left(\hat J_z^2
-\frac{5}{4}\, \hat 1 \right) \right] \quad , \\
H_1 &=& \frac{\hbar^2}{m_0} \, \sqrt{2}\,\, \tilde\gamma_2 \left( \{ k_z , k_+ \}
\{\hat J_z , \hat J_- \} + \{ k_z , k_- \} \{\hat J_z , \hat J_+ \} \right) \quad , \\
H_2 &=& \frac{\hbar^2}{2 m_0} \, \tilde\gamma_3 \left( k_+^2 \, \hat J_-^2 +
k_-^2 \, \hat J_+^2 \right) \quad .
\eaq
\endnumparts
Cartesian components of the spin-3/2 matrix vector are denoted by $\hat J_{x,y,z}$,
and we used the abbreviations $k_\pm = k_x \pm i k_y$, $\hat J_\pm = (\hat J_x
\pm i \hat J_y) / \sqrt{2}$, and $\{ A, B\} = (A B + B A)/2$. The constants
$\gamma_1$ and $\tilde\gamma_j$ are materials-dependent bandstructure
parameters~\cite{vurg:jap:01}. Note that the $\tilde\gamma_j$ depend also on the
quantum-well growth direction; their explicit expressions in terms of the standard
Luttinger parameters~\cite{luttham2,vurg:jap:01} $\gamma_2$ and $\gamma_3$
can be found, e.g., in Table~C.10 of Ref.~\cite{rolandbook}.

The dynamics of holes confined in a 2D quantum well is modeled by the Hamiltonian
$H_{\mathrm L} + V(z)$. In the following, we assume the external potential $V(z)$ to
be a hard-wall confinement of width $d$ and consider only its lowest size-quantised
orbital bound state. An effective Hamiltonian describing the 2D hole gas is then
obtained from \rf{eq:HL} by replacing $k_z \to \langle k_z \rangle = 0$ and $k_z^2
\to \langle k_z^2 \rangle = (\pi/d)^2$. Introducing the energy scale $E_0 = \pi^2
\hbar^2 \gamma_1 / (2 m_0 d^2)$ and measuring wave vectors in units of $\pi/d$,
the 2D hole-gas Hamiltonian is given by
\numparts
\baq{eq:HL2D}
H_{\mathrm L}^{\mathrm{(2D)}} &=& H_0^{\mathrm{(2D)}} + H_{\mathrm{mix}} \quad , \\
H_0^{\mathrm{(2D)}} &=& -E_0 \left\{ 1 - 2 \bar\gamma \left(\hat J_z^2 -\frac{5}{4}\,
\hat 1 \right) + \left[1 + \bar\gamma \left(\hat J_z^2 -\frac{5}{4}\, \hat 1 \right) \right]
\bar\vek{k}_\parallel^2 \right\} \quad , \\
H_{\mathrm{mix}} &=& E_0 \, \alpha \bar\gamma \left( \bar k_+^2 \, \hat J_-^2 +
\bar k_-^2 \, \hat J_+^2 \right) \quad .
\eaq
\endnumparts
Here $\bar k_{x,y} = k_{x,y} \, d/\pi$, $\bar\gamma = \tilde\gamma_1/\gamma_1$, and
$\alpha = \tilde\gamma_3/\tilde\gamma_1$. We are using the parameterisation in terms
of $\bar\gamma$ and $\alpha$ to be able to separately discuss the effects of HH-LH
splitting, which is embodied in $H_0^{\mathrm{(2D)}}$, and HH-LH mixing arising from
$H_{\mathrm{mix}}$.

Diagonalising $H_{\mathrm L}^{\mathrm{(2D)}}$ from \rf{eq:HL2D} yields in-plane
dispersion relations $E_j(\vek{k}_\parallel) = -E_0\,
\varepsilon_{\bar\vek{k}_\parallel}^{(j)}$ with $j=1,\dots, 4$ and
\be{eq:eigenVal}
\varepsilon_{\bar\vek{k}_\parallel}^{(j)}=1 + \bar{k}_x^2+\bar{k}_y^2 +
\sigma_j
\gb\sqrt{(\bar{k}_x^2+\bar{k}_y^2-2)^2+3\alpha^2(\bar{k}_x^2+\bar{k}_y^2)^2} \quad .
\ee
Here $\sigma_1=\sigma_2=-\sigma_3=-\sigma_4=1$. Using the result \rf{eq:eigenVal},
we obtain the two dimensionless Fermi wave vectors
\be{eq:FermiWave}
\bar{k}_{{\mathrm F}_{1,2}}=
\left[\frac{\varepsilon_{\mathrm F} -1 - 2 \gb^2 \mp \gb \sqrt{(\varepsilon_{\mathrm F}
-3)^2+ 3 \alpha^2 [(\varepsilon_{\mathrm F} - 1)^2 - 4\gb^2]}}{1 - \gb^2(1+3\alpha^2)}
\right]^{\frac{1}{2}} \quad ,
\ee
in terms of the dimensionless Fermi energy $\varepsilon_{\mathrm F}=-E_{\mathrm F}/
E_0$. The 2D-hole sheet density $n_{\mathrm{2D}}$ is related to the dimensionless
Fermi wave vectors according to
\be{eq:density}
n_{\mathrm{2D}} = \left(\frac{\pi}{d}\right)^2 \frac{\bar{k}_{{\mathrm F}_1}^2 \Theta(
\varepsilon_{\mathrm F}-[1+2\gb]) + \bar{k}_{{\mathrm F}_2}^2 \Theta(
\varepsilon_{\mathrm F}-[1-2\gb])}{2\pi}
\quad ,
\ee
where $\Theta(x)$ denotes the Heaviside step function.

The eigenvectors $|\chi_{\vek{k}_\parallel}^{(j)}\rangle$ corresponding to eigenvalues
$E_j(\vek{k}_\parallel)$ of $H_{\mathrm L}^{\mathrm{(2D)}}$ can be straightforwardly
determined. For $j=1,2$ and in the basis representation where $\hat J_z$ is diagonal,
we find
\be{eq:eigenVec}
%\hskip-2cm
|\chi_{\vek{k}_\parallel}^{(1)} \rangle=
\left(\begin{array}{cccc}
0 \\[2mm]
\frac{(s-t)(\bar{k}_x-i \bar{k}_y)^2}{\bar\vek{k}_\parallel^2
\sqrt{(s-t)^2+3\alpha^2\bar\vek{k}_\parallel^4}}\\[2mm]
0\\[2mm]
\frac{\sqrt{3}\alpha\bar\vek{k}_\parallel^2}{\sqrt{(s-t)^2+3\alpha^2
\bar\vek{k}_\parallel^4}}
\end{array}\right),\qquad
|\chi_{\vek{k}_\parallel}^{(2)} \rangle=
\left(\begin{array}{cccc}
\frac{(-s-t)(\bar{k}_x-i \bar{k}_y)^2}{\bar\vek{k}_\parallel^2\sqrt{(-s-t)^2+3
\alpha^2\bar\vek{k}_\parallel^4}}\\[2mm]
0 \\[2mm]
\frac{\sqrt{3}\alpha\bar\vek{k}_\parallel^2}{\sqrt{(-s-t)^2+3
\alpha^2 \bar\vek{k}_\parallel^4}}\\[2mm]
0 
\end{array}\right),
\ee
with $s \equiv \bar\vek{k}_\parallel^2-2$ and $t\equiv \sqrt{s^2+3\alpha^2 \bar
\vek{k}_\parallel^4}$. The remaining eigenspinors $|\chi_{\vek{k}_\parallel}^{(3)}
\rangle$ and $|\chi_{\vek{k}_\parallel}^{(4)}\rangle$ are obtained by changing $t\to-t$
in $|\chi_{\vek{k}_\parallel}^{(1)}\rangle$ and $|\chi_{\vek{k}_\parallel}^{(2)}\rangle$,
respectively. As the scalar products $\langle \chi_{\vek{k}_\parallel}^{(j)}|
\chi_{\vek{k}_\parallel+\vek{q}}^{(l)} \rangle$ enter in the calculation of the
Lindhard function, we briefly discuss their relevant properties. The moduli $|\langle
\chi_{\vek{k}_\parallel}^{(j)}|\chi_{\vek{k}_\parallel+\vek{q}}^{(j)} \rangle|$ are found
to be equal for all $j=1,\dots,4$. Also, $|\langle \chi_{\vek{k}_\parallel}^{(j)}|
\chi_{\vek{k}_\parallel+\vek{q}}^{(l)} \rangle|$ are pairwise the same for $(j,l)=(1,3)$
and $(3,1)$; and $(j,l)=(2,4)$ and  $(4,2)$. These relations can be verified using the
explicit form of the column vectors in \rf{eq:eigenVec} together with the fact that the
eigenspinors satisfy orthonormality relations.

%%%%%%%%%%%%%%%%%%%%%%%%%%%%%%%%%%%%%% 
\section{Lindhard function of a 2D hole gas: General expression and special
cases}
\label{sec:lindhard}
%%%%%%%%%%%%%%%%%%%%%%%%%%%%%%%%%%%%%% 

The general definition~\cite{GiulianiVignale} of the Lindhard function, specialised
to a 2D system, reads
\be{eq:respondsF}
\hskip-1cm
\chi(\omega,\vek{q})=\lim_{\delta\to 0}
\sum^4_{j,l=1}\int \frac{{\mathrm d}^2 k_\parallel}{(2\pi)^2}
|\langle \chi_{\vek{k}_\parallel}^{(j)}| \chi_{\vek{k}_\parallel+\vek{q}}^{(l)} \rangle|^2 
\frac{n_{\mathrm F} [ E_j(\vek{k}_\parallel)] - n_{\mathrm F} [ E_l (\vek{k}_\parallel+
\vek{q})]}{\hbar\omega + i \delta + E_j(\vek{k}_\parallel) - E_l (\vek{k}_\parallel+\vek{q})}
\quad ,
\ee
with $n_F(E)$ denoting the Fermi-Dirac distribution function. The expression given in
\rf{eq:respondsF} can be simplified by using a polar-coordinate representation where
$k_x = k_\parallel\cos\phi$ and $k_y=k_\parallel\sin\phi$ and performing a change of
variables in the terms involving $n_{\mathrm F}[E_j(\vek{k}_\parallel+\vek{q})]$ such
that $\vek{k}_\parallel\to\vek{k}_\parallel-\vek{q}$ and $\phi\to \phi+\pi$, which leaves
the energy difference and the spinor overlap invariant. Using also the description in terms
of dimensionless quantities and specialising to the zero-temperature limit, the Lindhard
function can be expressed as $\chi(\omega,\vek{q})= -(2 m_0 / \hbar^2 \gamma_1)\,
\bar\chi(\bar\omega, \bar q)$, with
\be{eq:lindhard}
\hskip-1cm
\bar\chi(\bar{\omega},\bar{q})=
\lim_{\delta\to 0} \sum_{\eta=\pm 1} \sum^4_{j,l=1}\int_0^{\bar{K}_{{\mathrm F}_j}}
\bar{k}_\parallel \, {\mathrm d}\bar{k}_\parallel \int_0^{2\pi} \frac{{\mathrm d}\phi}{(2\pi)^2}
\frac{|\langle \chi_{\bar{\vek{k}}_\parallel}^{(j)} | \chi_{\bar{\vek{k}}_\parallel+
\bar{\vek{q}}}^{(l)} \rangle|^2}{\eta(\bar{\omega}+i\delta)+
\varepsilon_{\bar{\vek{k}}_\parallel}^{(j)} - \varepsilon_{\bar{\vek{k}}_\parallel+
\bar{\vek{q}}}^{(l)}} \quad .
\ee
We use a notation where $\bar{K}_{{\mathrm F}_1}=\bar{K}_{{\mathrm F}_2} \equiv
\bar{k}_{{\mathrm F}_1}$, $\bar{K}_{{\mathrm F}_3}=\bar{K}_{{\mathrm F}_4} \equiv
\bar{k}_{{\mathrm F}_2}$, and $\bar{\omega}=\hbar\omega/E_0$. Note that, because the
Luttinger Hamiltonian in axial approximation exhibits rotational invariance of in-plane hole
motion, the Lindhard function depends on wave vector $\vek{q}$ only via its (dimensionless)
magnitude $\bar q$. Also, within our effective 2D description, $\chi(\omega,\vek{q})$ is
independent of the quantum-well width $d$ and inversely proportional to $\gamma_1$.

In the following, we consider the static limit, which is obtained by setting $\bar{\omega}=0$.
Specialising further to certain limiting situations, we can find analytical expressions for the Lindhard function. For example, for the case of vanishing HH-LH mixing obtained by letting
$\alpha\to 0$, the matrix $| \langle \chi_{\vek{k}_\parallel}^{(j)}|
\chi_{\vek{k}_\parallel+\vek{q}}^{(l)} \rangle|^2$ of modulus-squared scalar products
reduces to the unity matrix, and the simple analytical expression
\be{eq:LFnomix}
\left. \bar\chi(0,\bar{q}) \right|_{\alpha=0} =
\frac{-1}{2\pi\bar{q}} \sum_{j=1}^2 \left\{
\frac{\Theta(\bar k_{{\mathrm{F}}_j})}{1-\sigma_j \gb} \left[ \bar{q} - \sqrt{\bar{q}^2 -
4\bar{k}^2_{{\mathrm F}_j}}\,  \Theta \left(\frac{\bar{q}}{2\bar{k}_{{\mathrm F}_j}} - 1
\right)\right]\right\} \,\,\, ,
\ee
is found, where $\sigma_1=-\sigma_2 = 1$. Inspection of the result \rf{eq:LFnomix} shows
that, with only HH-LH splitting included, the static Lindhard function comprises two separate
HH and LH contributions, each being the standard 2D-electron-gas
expression~\cite{GiulianiVignale} with Fermi wave vector and effective mass adjusted to
the respective HH and LH values. On the other hand, taking the limit $\bar{\vek{q}}\to 0$
in \rf{eq:lindhard}, the matrix of modulus-squared spinor overlaps again becomes the unity
matrix, and we find an analytical result for the (dimensionless) density of states at the
Fermi energy,
\be{eq:Lindhardq0}
\lim_{\bar{q}\to 0} \bar\chi(0,\bar{q})=
\frac{-1}{2\pi}\sum_{j=1}^2 \Theta(\bar k_{{\mathrm{F}}_j}) \,
\left| 1- \sigma_j \gb \, 
\frac{2-\bar k^2_{{\mathrm F}_j}(1+3\alpha^2)}{\sqrt{(2-\bar k^2_{{\mathrm F}_j})^2
+3\alpha^2 \bar k^4_{{\mathrm F}_j}}} \right|^{-1} \quad .
\ee
Thus we see that one effect of HH-LH mixing is to introduce an energy (and concomitant
density) dependence into the density of states of 2D holes.

%%%%%%%%%%%%%%%%%%%%%%%%%%%%%%%%%%%%%%%%%%
\section{Static polarizability of 2D holes: Numerical method and results}
\label{sec:results}
%%%%%%%%%%%%%%%%%%%%%%%%%%%%%%%%%%%%%%%%%%

With analytical expressions unavailable for the Lindhard function \rf{eq:lindhard} in
the more general case with both $\vek{q}$ and $\alpha$ finite, we have to resort to
numerical calculations to investigate in greater detail how HH-LH mixing affects the
static polarizability $\bar\chi(0,\bar{q})$. Note that the latter is an entirely real-valued
function. The procedure for its numerical calculation is explained in the following
Subsection, and our results are given thereafter.

\subsection{Brief outline of the calculational method}

For $\bar\omega=0$, the integrand of \rf{eq:lindhard} has poles whenever the energy
difference in the denominator vanishes. These poles are regularised by the parameter
$\delta$, which needs to be set to zero after performing the integrations. We calculate
these integrals numerically, taking special care in the regions close to values of the
integration variables corresponding to a vanishing denominator. To identify the pole
structure of the Lindhard function, we write the inverse of the energy difference as
\begin{eqnarray}
\nonumber
\left( \varepsilon_{\bar{\vek{k}}_\parallel}^{(j)} - \varepsilon_{\bar{\vek{k}}_\parallel+ 
\bar{\vek{q}}}^{(l)} \right)^{-1} & = &(\delta_{j,1}+\delta_{j,2})(\delta_{l,1}+\delta_{l,2})
\frac{a_1-b}{a_1^2-b^2}\\ 
& +& \nonumber(\delta_{j,1}+\delta_{j,2})(\delta_{l,3}+\delta_{l,4}) \frac{a_1+b}{a_1^2-b^2}\\
&+ &\nonumber (\delta_{j,3}+\delta_{j,4})(\delta_{l,1}+\delta_{l,2}) \frac{a_2-b}{a_2^2-b^2}\\ 
\label{eq:Polestruc}
&+ &(\delta_{j,3}+\delta_{j,4})(\delta_{l,3}+\delta_{l,4}) \frac{a_2+b}{a_2^2-b^2},
\end{eqnarray}
where $\delta_{j,l}$ denotes Kronecker's delta symbol. The quantities appearing in
(\ref{eq:Polestruc}) are
\numparts
\baq{eq:short}
&&a_{1,2}=\mp\gb\sqrt{4-4\bar k_\parallel^2+(1+3\alpha^2)\bar k_\parallel^4}+\bar q^2+
\bar k_\parallel \bar q\cos\phi\\[2mm]
\hskip-2cm
&&b =\gb\sqrt{(-2+\bar k_\parallel^2+ \bar q^2+2\bar k_\parallel \bar q \cos\phi)^2
+3\alpha^2( \bar k_\parallel^2+ \bar q^2+2\bar k_\parallel \bar q\cos\phi)^2}~.
\eaq
\endnumparts
The denominators in (\ref{eq:Polestruc}) can be written as
\be{eq:denomiPole}
\hskip-2cm
\frac{1}{a_{1,2}^2-b^2}=\frac{1}{4 \bar k_\parallel^2 \bar q^2[1-(1+3\alpha^2)\gb^2]}
\frac{1}{X_{1,2}-Y_{1,2}}
\left(\frac{1}{\cos\phi-X_{1,2}}-\frac{1}{\cos\phi-Y_{1,2}}\right)~,
\ee
with the positions of the poles given by 
\baq{eq:Poleposi}
\hskip-2cm
&&X_{1,2}=-\frac{\bar q}{2\bar k_\parallel}~,\nonumber\\[2mm]
\hskip-2cm
&&Y_{1,2}=\frac{\pm2\gb\sqrt{4-4\bar k_\parallel^2+(1+3\alpha^2) \bar k_\parallel^4}
-\bar q^2+\gb^2 [(2+6\alpha^2) \bar k_\parallel^2-4+\bar q^2+3\alpha^2 \bar q^2]}
{2\bar k_\parallel \bar q[1-(1+3\alpha^2)\gb^2]}~.
\eaq
As can be seen from \rf{eq:denomiPole}, poles are encountered in the 
integration over $\phi$ when $|X_{1,2}|,|Y_{1,2}|\leq 1$. 
We have employed a Cauchy principle-value integration to regularise the
Lindhard function in the vicinity of the poles specified in \rf{eq:Poleposi}.
 
\subsection{Results for model parameters applying to an [001] quantum well
in GaAs}

High-quality 2D hole gases have recently been fabricated from [001]-grown
GaAs heterostructures~\cite{ensslin:apl:04,manfra:apl:05,weg:apl:05b}. To
obtain results applicable to these systems, we use the appropriate model
parameters $\gb = 0.31$ and $\alpha = 1.2$. Results for this configuration
are presented below. For comparison and to clearly show the impact of
HH-LH mixing, we also show results for the static polarizability when
$\alpha = 1$, $0.5$, and $0$. In all these cases, we limit ourselves to the
low-density regime where only the highest, HH-like, 2D subband is occupied.
The reason for this caution is the fact that, within our model using a hard-wall
confinement, the spectrum of all other than the highest subband poorly
matches that of the real GaAs sample. To be specific, we choose
$\bar{n}=0.0608$. Recalling that the 2D-hole sheet density is related to the
dimensionless density by $n=(\pi/d)^2\bar{n}$, this value would correspond
to a density of $n=1.5\times10^{15}$~m${}^{-2}$ in a $20$-nm quantum well.

\begin{figure}[t]
\begin{indented}
\item[] \hspace{-1cm}\includegraphics[width=8cm]{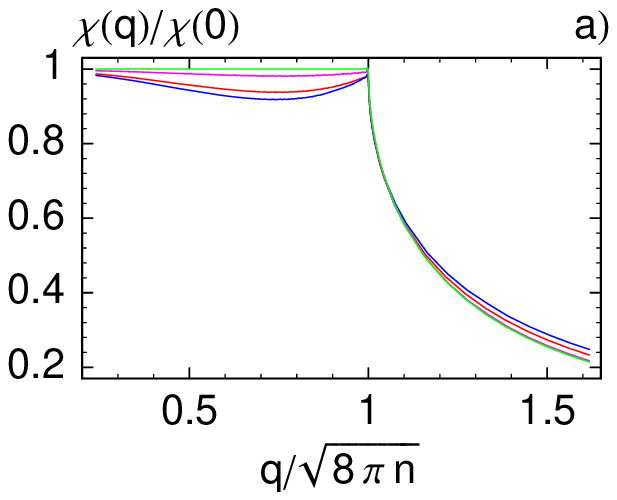}\hspace{-1cm}
\includegraphics[width=8cm]{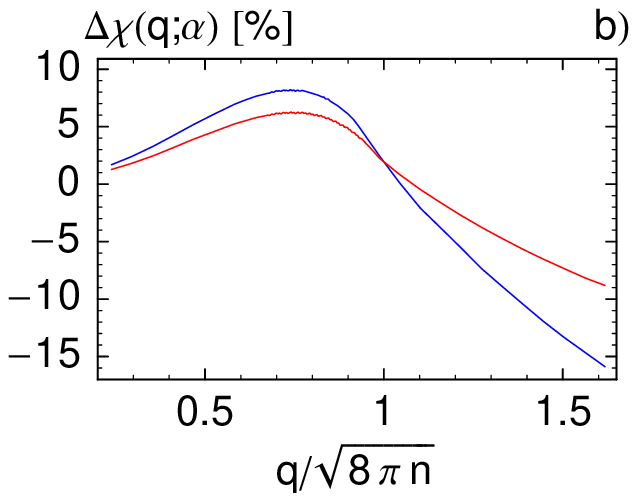}
\caption{(a)~Normalised Lindhard function $\chi(q)$ for the static limit and
(b)~the quantity $\Delta\chi(q;\alpha)$ that measures the impact of HH-LH
mixing (see text), plotted as a function of wave-vector magnitude $q$. The
blue curve is for $\bar{\gamma}=0.31$ and $\alpha=1.2$, which are parameter
values applying to an [001]-grown heterostructure in GaAs. For comparison,
results are also shown for $\alpha=1$ (red curve), $0.5$ (magenta curve),
and $0$ (green curve). For all cases, a value $\bar{n}=0.0608$ for the
dimensionless 2D hole density was used.
\label{figure1}}
\end{indented}
\end{figure}
To avoid cluttering our notation, we suppress the zero-frequency argument in
the formal expression of the static Lindhard function from now on: $\chi(0, \vek{q})
\equiv \chi(q)$. In Figure~\ref{figure1}(a), we plot $\chi(q)/\chi(0)$ as a function of
$q/\sqrt{8\pi n}$, for different values of the parameter $\alpha$ that quantifies the
HH-LH mixing. It is apparent that a finite $\alpha$ leads to a significant suppression
of $\chi(q)$ below the constant-plateau value usually associated with 2D
systems~\cite{GiulianiVignale} for $q < \sqrt{8\pi n}$. To make the
impact of HH-LH mixing quantitatively explicit, we define the variable
\be{eq:Delta}
\Delta\chi(q;\alpha) = 1 - 
\left. \frac{\chi(q)}{\chi(0)}\right|_\alpha \left[ \left. \frac{\tilde\chi(q)}{\tilde\chi(0)}
\right|_\alpha \right]^{-1} \quad ,
\ee
where $\tilde\chi(q)$ is the analytical result \rf{eq:LFnomix} obtained for the
limit $\alpha=0$ but with Fermi wave vectors adjusted to coincide with those found
in the case of the finite $\alpha$ under consideration. Thus the function
$\Delta\chi(q;\alpha)$ measures the relative change exhibited in the normalised
static polarizability that is due to a finite $\alpha$ but goes beyond a simple
renormalisation of Fermi wave vectors.\footnote{In the low-density limit considered
here, there is only one Fermi wave vector whose magnitude is the same
for all values of $\alpha$, and $\tilde\chi(q)$ actually coincides with
$\chi(q)|_{\alpha=0}$. However, as we will see further below,
$\tilde\chi(q)\ne\chi(q)|_{\alpha=0}$ in the more general case when both the
HH and LH subbands arising from the lowest 2D orbital bound state are occupied
and, thus, two Fermi wave vectors exist.}
In Figure~\ref{figure1}(b), we show $\Delta\chi(q;\alpha)$ for $\alpha=1$ (red curve)
and $\alpha=1.2$ (blue curve). It shows a strong variation as a function of
$q/\sqrt{8\pi n}$ and reaches the 10$\%$ level.

\subsection{Results for a model semiconductor: High- and low-density regimes}

For high-enough 2D densities, holes will occupy both the HH-like and LH-like
subbands arising from the lowest-energy orbital bound state in the quantum well.
It can be expected that this high-density regime is qualitatively different from the
situation at low density where only the highest (HH-like) 2D subband is occupied.
To treat the case of high density consistently within our adopted model, it needs
to be ensured that the LH-like subband arising from the lowest-energy orbital
bound state is still higher in energy than the HH-like subband associated
with the next orbital-bound-state level. For a hard-wall confinement considered
here, a system with $\gb=0.2$ satisfies that condition. Although this value does
not directly correspond to a specific semiconductor material, we use it to illustrate
the generically different impact of HH-LH mixing in the low and high-density regimes,
respectively.

\begin{figure}[t]
\begin{indented}
\item[] \includegraphics[width=9cm]{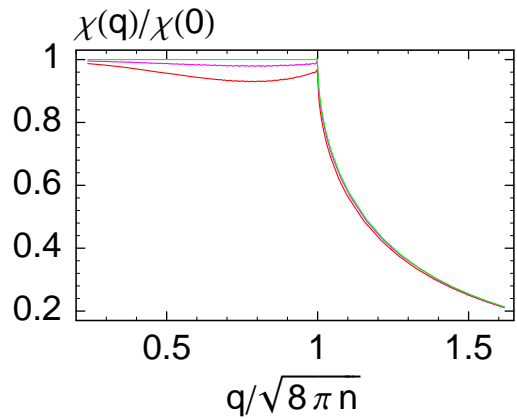}
\caption{Normalised Lindhard function $\chi(q)$ for the low-density regime of a
fictitious semiconductor material with $\bar{\gamma}=0.2$ and (dimensionless)
density $\bar{n}=0.0608$. The red, magenta, and green curves correspond to
$\alpha=1$, $0.5$, and $0$, respectively.
\label{figure2}}
\end{indented}
\end{figure}
To provide a clear benchmark for comparing high and low-density regimes, we start
by presenting the result for the low density case ($\bar{n}=0.0608$, same value
as used in the calculations for Figure~\ref{figure1}) in Figure~\ref{figure2}. The obtained
curves look qualitatively similar to those found for the low-density regime in GaAs
(different $\gb$, shown in Figure~\ref{figure1}), but the quantitative level of suppression
below the plateau value obtained in the limit of vanishing $\alpha$ is different here.

\begin{figure}[t]
\begin{indented}
\item[] \hspace{-1cm}\includegraphics[width=8cm]{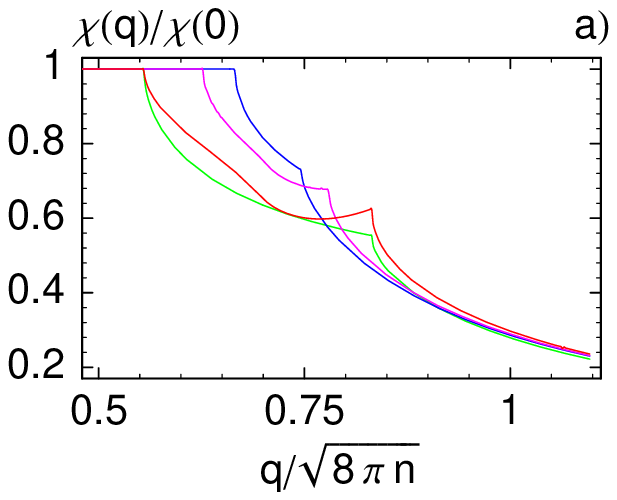}\hspace{-1cm}
\includegraphics[width=8cm]{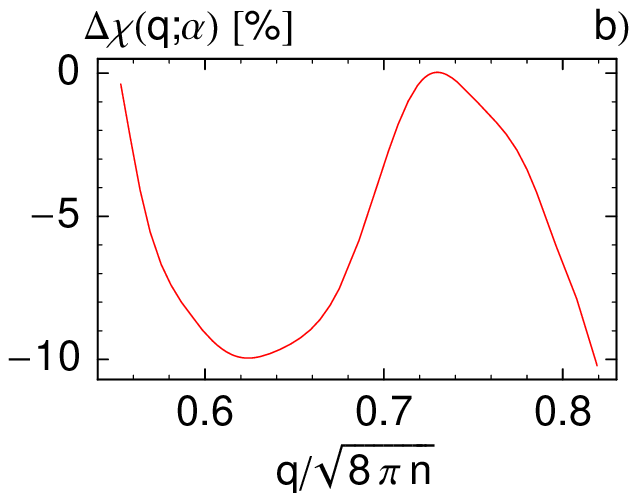}
\caption{(a)~Normalised Lindhard function $\chi(q)$ for the static limit and
(b)~the quantity $\Delta\chi(q;\alpha)$ that measures the impact of HH-LH
mixing (see text), for the case with $\bar{\gamma}=0.2$ and (dimensionless)
density $\bar{n}=0.4055$, corresponding to the high-density regime with both
HH-like and LH-like 2D subbands occupied. The red, magenta, and blue curves
are obtained for $\alpha=1$, $0.5$, and $0$, respectively. For comparison, we also
plot (as the green curve) the analytical result \rf{eq:LFnomix} with Fermi-wave-vector
values adjusted to coincide with the $\alpha=1$ case. Therefore, the red and green
curves in panel~(a) exhibit kink-like features at the same values of $q$, and
deviations between the two illustrate the effect of HH-LH mixing beyond a simple
renormalisation of the Fermi wave vectors.
\label{figure3}}
\end{indented}
\end{figure}
As an illustration of the high-density regime, we present results for $\bar{n}=0.4055$,
which would correspond to $n=10^{16}$~m${}^{-2}$ in a $20$-nm quantum well. The
normalised Lindhard function for this case is plotted in Figure~\ref{figure3}(a). Note that, for
the three values of $\alpha$ for which results are presented, the two Fermi wave vectors
are different, see \rf{eq:FermiWave}. As a result, the sharp features arising in the Lindhard
function from poles at $-q/2k_{F_{1,2}}$ appear at different values of $q$ in each curve. 
In contrast to the low-density case, a plateau is again exhibited in the static Lindhard
function (for $q\le 2 k_{{\mathrm F}_1}$). To illustrate effects due to HH-LH mixing beyond
a simple renormalisation of the two Fermi wave vectors, we also show (as the green curve)
the analytical result \rf{eq:LFnomix} for the Lindhard function in the limit $\alpha=0$ but
with values for the Fermi wave vectors taken from the case $\alpha=1$. The latter result
corresponds to that expected for two independent 2D hole gases with different Fermi wave
vectors. Thus, any deviation between the red and green curves is entirely due to the mixed
HH-LH character of 2D hole states. Figure~\ref{figure3}(b) shows the corresponding
difference function $\Delta\chi(q;\alpha)$ for $\alpha=1$. Again, HH-LH mixing appears to
cause differences around the 10$\%$-level.

%%%%%%%%%%%%%%%%%%%%%%%%%%%%%%%%%%%%%%%%%%
\section{Summary and Conclusions}
\label{sec:concl}
%%%%%%%%%%%%%%%%%%%%%%%%%%%%%%%%%%%%%%%%%%

We have calculated the density-density response function of a homogeneous 2D hole
gas in the static limit, based on the Luttinger-model description of the upper-most
valence band within the axial approximation. While this approach neglects the warping of
energy dispersions (and, hence, Fermi surfaces) for the holes' in-plane motion due to
the cubic crystal symmetry, it already captures essential new features arising from the
peculiar valence-band properties. We furthermore focused only on the lowest orbital
bound state in a symmetric quantum well defined by a hard-wall confinement. HH-LH
splitting gives rise to the existence of two energetically separated 2D hole subbands,
one (at higher energy) mostly HH-like and the other of mostly LH character. However,
except for states with zero in-plane kinetic energy, HH and LH amplitudes are mixed,
and we have elucidated how this mixing gives rise to marked changes in the shape and
magnitude of the static density response function. New analytical results are derived
for the limit $q\to 0$, but the case with finite $q$ and HH-LH mixing included could
only be treated numerically.

Both for practical and theoretical reasons, it makes sense to distinguish two basic
situations. One corresponds to the low-density regime where only the highest (mostly
HH-like) 2D subband is occupied. In this limit, our model can be expected to describe
real semiconductor heterostructures  quite accurately, even quantitatively, if adequate
band-structure parameters are used as input to our calculations. As it turns out, even
though holes are present only in the HH-like 2D subband, HH-LH mixing importantly
affects the static density response. In particular, the response function is suppressed
below the plateau exhibited by the standard 2D-electron-gas result~\cite{GiulianiVignale},
as illustrated in Figures~\ref{figure1}(a) and \ref{figure2}. In the high-density regime,
where both the HH-like and LH-like 2D subbands associated with the lowest-energy
quantum-well bound state are occupied, the density response differs from that expected
for two independent 2D hole gases. Thus HH-LH mixing is shown to influence the
Lindhard function beyond a trivial renormalisation of Fermi-wave-vector magnitudes.
We have defined, and calculated, the quantity $\Delta\chi(q;\alpha)$ to make the
nontrivial effects arising from HH-LH mixing quantitatively explicit. For the parameters
considered, relative changes on the order of 10\% are seen.

In this work, we employed the approximation of hard-wall (infinite-height) quantum-well
barriers. In the more realistic case of finite-height barriers, the HH and LH bound-state
wave functions will both penetrate into the barrier regions. Due to HH-LH splitting, the
range of penetration will be higher for LH states, i.e., the effective quantum-well width
will be larger for LHs. This subtle difference between HH and LH bound states scales
with the parameter $\bar\gamma$ and can be expected to result in corrections of order
$\alpha\bar\gamma^2$ to effects due to HH-LH mixing.

The qualitative and quantitative impact of HH-LH mixing can be expected to affect
physical properties of 2D hole gases in an important way and, thus, render their
behaviour quite different from that exhibited by 2D conduction-electron systems.
Examples for physical observables that are affected include the shape and range
of Friedel oscillations exhibited by 2D hole gases in response to impurity charges
present, e.g., in the doping layer of modulation-doped heterostructures. HH-LH
mixing should then also influence the 2D-hole-mediated Ruderman-Kittel-Kasuya-Yosida
(RKKY) interaction between magnetic impurities. A detailed investigation of this effect could
shed new light on how to tailor the ferromagnetic properties of 2D
diluted-magnetic-semiconductor
heterostructures~\cite{haury:prl:97,wojto:apl:03,nazmul:prb:03,nazmul:prl:05,weiss:prb:09}.

% \ack
% Any acknowledgements go here.

%%%%%%%%%%%%%%%%%%%%%%%%%%%%%%%%%%%%%%%%%%
\section*{References}
%%%%%%%%%%%%%%%%%%%%%%%%%%%%%%%%%%%%%%%%%%

% \bibliographystyle{iopart-num}
% \bibliography{general,spintronics,spinorbit}

\begin{thebibliography}{10}
\expandafter\ifx\csname url\endcsname\relax
  \def\url#1{{\tt #1}}\fi
\expandafter\ifx\csname urlprefix\endcsname\relax\def\urlprefix{URL }\fi
\providecommand{\eprint}[2][]{\url{#2}}

\bibitem{Ziman}
Ziman J~M 1972 {\em Theory of Solids\/} 2nd ed (Cambridge, UK: Cambridge U
  Press)

\bibitem{GiulianiVignale}
Giuliani G and Vignale G 2005 {\em Quantum Theory of the Electron Liquid\/}
  (Cambridge, UK: Cambridge U Press)

\bibitem{hetero1}
Ando T, Fowler A~B and Stern F 1982 {\em Rev. Mod. Phys.\/} {\bf 54} 437

\bibitem{raikh:prb:99}
Chen G~H and Raikh M~E 1999 {\em Phys. Rev. B\/} {\bf 59} 5090

\bibitem{wang:prb:05}
Wang X~F 2005 {\em Phys. Rev. B\/} {\bf 72} 085317

\bibitem{mishch:prl:06}
Farid A~K and Mishchenko E~G 2006 {\em Phys. Rev. Lett.\/} {\bf 97} 096604

\bibitem{pletyu:prb:06}
Pletyukhov M and Gritsev V 2006 {\em Phys. Rev. B\/} {\bf 74} 045307

\bibitem{fabian:prb:09}
Badalyan S~M, Matos-Abiague A, Vignale G and Fabian J 2009 {\em Phys. Rev. B\/}
  {\bf 79} 205305

\bibitem{qmc}
Ambrosetti A, Pederiva F, Lipparini E and Gandolfi S 2009 {\em Phys. Rev. B\/}
  {\bf 80} 125306

\bibitem{zutic:rmp:04}
Zuti\'{c} I, Fabian J and Sarma S~D 2004 {\em Rev. Mod. Phys.\/} {\bf 76}
  323

\bibitem{flatte:natphys:06}
Awschalom D~D and Flatt{\'e} M~E 2006 {\em Nat. Phys.\/} {\bf 3} 153

\bibitem{YuCardona}
Yu P~Y and Cardona M 1999 {\em Fundamentals of Semiconductors\/} 2nd ed
  (Berlin: Springer)

\bibitem{schlie:prb:06b}
Schliemann J 2006 {\em Phys. Rev. B\/} {\bf 74} 045214

\bibitem{galitski:prb:06}
Stanescu T~D and Galitski V 2006 {\em Phys. Rev. B\/} {\bf 74} 205331

\bibitem{schlie:unpub:10}
Schliemann J The dielectric function of the semiconductor hole gas preprint
  arXiv:1003.4820

\bibitem{ensslin:apl:04}
Grbi\'{c} B, Ellenberger C, Ihn T, Ensslin K, Reuter D and Wieck A~D 2004 {\em
  Appl. Phys. Lett.\/} {\bf 85} 2277

\bibitem{manfra:apl:05}
Manfra M~J, Pfeiffer L~N, West K~W, de~Picciotto R and Baldwin K~W 2005 {\em
  Appl. Phys. Lett.\/} {\bf 86} 162106

\bibitem{matt:apl:05}
Fischer F, Schuh D, Bichler M, Abstreiter G, Grayson M and Neumaier K 2005 {\em
  Appl. Phys. Lett.\/} {\bf 86} 192106

\bibitem{weg:apl:05a}
Schmult S, Gerl C, Wurstbauer U, Mitzkus C and Wegscheider W 2005 {\em Appl.
  Phys. Lett.\/} {\bf 86} 202105

\bibitem{weg:apl:05b}
Gerl C, Schmult S, Tranitz H~P, Mitzkus C and Wegscheider W 2005 {\em Appl.
  Phys. Lett.\/} {\bf 86} 252105

\bibitem{clarke:jap:06}
Clarke W~R, Micolich A~P, Hamilton A~R, Simmons M~Y, Muraki K and Hirayama Y
  2006 {\em J. Appl. Phys.\/} {\bf 99} 023707

\bibitem{rolandbook}
Winkler R 2003 {\em Spin-Orbit Coupling Effects in Two-Dimensional Electron and
  Hole Systems\/} (Berlin: Springer)

\bibitem{bastardrev}
Bastard G, Brum J~A and Ferreira R 1991 {\em Solid State Physics\/} vol~44
ed H Ehrenreich and D Turnbull (San Diego: Academic Press) pp 229--415

\bibitem{luttham1}
Luttinger J~M and Kohn W 1955 {\em Phys. Rev.\/} {\bf 97} 869

\bibitem{luttham2}
Luttinger J~M 1956 {\em Phys. Rev.\/} {\bf 102} 1030

\bibitem{hensel:prb:74}
Suzuki K and Hensel J~C 1974 {\em Phys. Rev. B\/} {\bf 9} 4184

\bibitem{treb:prb:79}
Trebin H~R, R\"ossler U and Ranvaud R 1979 {\em Phys. Rev. B\/} {\bf 20}
  686

\bibitem{roessler:prb:91}
Mayer H and R\"ossler U 1991 {\em Phys. Rev. B\/} {\bf 44} 9048

\bibitem{vurg:jap:01}
Vurgaftman I, Meyer J~R and Ram-Mohan L~R 2001 {\em J. Appl. Phys.\/} {\bf 89}
  5815

\bibitem{haury:prl:97}
Haury A, Wasiela A, Arnoult A, Cibert J, Tatarenko S, Dietl T and
  Merle~d'Aubign\'e Y 1997 {\em Phys. Rev. Lett.\/} {\bf 79} 511

\bibitem{wojto:apl:03}
Wojtowicz T, Lim W~L, Liu X, Dobrowolska M, Furdyna J~K, Yu K~M, Walukiewicz W,
  Vurgaftman I and Meyer J~R 2003 {\em Appl. Phys. Lett.\/} {\bf 83} 4220

\bibitem{nazmul:prb:03}
Nazmul A~M, Sugahara S and Tanaka M 2003 {\em Phys. Rev. B\/} {\bf 67} 241308

\bibitem{nazmul:prl:05}
Nazmul A~M, Amemiya T, Shuto Y, Sugahara S and Tanaka M 2005 {\em Phys. Rev.
  Lett.\/} {\bf 95} 017201

\bibitem{weiss:prb:09}
Wurstbauer U and Wegscheider W 2009 {\em Phys. Rev. B\/} {\bf 79} 155444

\end{thebibliography}

\providecommand{\newblock}{}

\end{document}